\documentclass[11pt,a4paper,twoside]{article}

\usepackage[margin=1in]{geometry}
\usepackage{graphicx}
\usepackage{amsmath,amssymb,bm}
\usepackage{cite}
\usepackage{hyperref}
\usepackage{fancyhdr}
\usepackage{float}
\usepackage{subcaption}
\usepackage{siunitx}
\usepackage{xcolor}
\usepackage{booktabs}

\DeclareSIUnit\cps{cps}

\hypersetup{
    colorlinks=true,
    linkcolor=blue,
    citecolor=blue,
    urlcolor=blue
}

\newcommand{\ket}[1]{\left|#1\right\rangle}
\newcommand{\bra}[1]{\left\langle#1\right|}
\newcommand{\Tr}{\mathrm{Tr}}

\newcommand{\figinclude}[2][]{\includegraphics[#1]{#2}}

\title{
Broadband Telecom Entanglement from an All-Fiber Type-0 Sagnac Source
}

\author{
\resizebox{\dimexpr\textwidth-2\tabcolsep\relax}{!}{%
Alessandro Cestelli$^{1,*}$, Spyridon Grountas$^{1}$, Scott Mc Haffie$^{1}$, Jun Gao$^{1}$, Val Zwiller$^{2,3}$, Ali Elshaari$^{1}$%
}\\[0.4cm]
\small $^{1}$KTH Royal Institute of Technology, Stockholm, Sweden\\
\small $^{2}$Rise Research Institutes of Sweden, Stockholm, Sweden\\
\small $^{3}$Quantum Scopes AB, Stockholm, Sweden\\[0.15cm]
\small $^{*}$Corresponding authors: cestelli@kth.se
}

\date{}

\begin{document}

\raggedbottom
\emergencystretch=1em
\clubpenalty=10000
\widowpenalty=10000

\maketitle

% ============================================================
\begin{abstract}
We demonstrate a compact, practical all-fiber source of non-degenerate
telecom-band polarization-entangled photon pairs assembled from
off-the-shelf components. The compact fiber-integrated architecture is
designed to be affordable, straightforward to assemble and align, and
compatible with existing optical-fiber infrastructure. Photon pairs are generated through type-0
spontaneous parametric down-conversion in a pigtailed periodically poled
lithium-niobate waveguide embedded in a Sagnac interferometer. At
\SI{0.5}{\milli\watt} pump power, quantum-state tomography yields a
Bell-state fidelity $F_{\Phi^-}=0.850$, purity $\mathcal{P}=0.836$, and
concurrence $C=0.782$. A separate CHSH measurement gives
$S=2.3301\pm0.0075$. The measured Bell-state fidelity is sufficient for
the reconstructed polarization state to support qubit teleportation
above the classical fidelity limit under ideal protocol operations.
Analysis of the density matrix shows that approximately $96.7\%$ of
the population remains in the desired $\ket{HH}$--$\ket{VV}$ subspace,
with a normalized coherence of approximately $0.86$. A residual
relative-phase offset indicates that local phase compensation could
increase the Bell-state fidelity to approximately $0.8865$. We further characterize the brightness--quality trade-off and obtain
a corrected pair-rate slope of
\SI{0.918(40)}{\mega\cps\per\milli\watt}. The
broadband emission and fiber interfaces offer a route to multiuser
entanglement distribution through off-the-shelf multichannel
wavelength-division multiplexers with passbands matched to the signal
and idler wavelengths.
\end{abstract}

% ============================================================
\section{Introduction}
% ============================================================

Entangled photon pairs are a central resource for quantum
communication, photonic quantum networking,
and fundamental tests of nonclassical correlations
\cite{Ekert1991,Wehner2018,Clauser1969,Weihs1998,Kwiat1995}. Practical
sources must combine high entanglement quality and brightness with
mechanical stability and compatibility with existing fiber
infrastructure, motivating guided-wave and fiber-integrated sources at
telecom wavelengths \cite{Tanzilli2002,Tanzilli2012,Vergyris2017}.
Polarization entanglement is particularly attractive because of its
simple preparation and measurement, although propagation through fiber
systems can introduce correctable polarization rotations, while
unresolved spectral--temporal distinguishability, including that induced
by birefringent delays, can reduce polarization coherence.

Sagnac interferometers provide a robust geometry for generating
polarization entanglement because the two pair-generation amplitudes
propagate through largely common optical paths
\cite{Kim2006,Fedrizzi2007,Kim2019}. Incorporating periodically poled
waveguides into the loop combines high nonlinear efficiency with fiber
compatibility and compact construction. For broadband sources, however,
the quality of the polarization state also depends on the
spectral--temporal overlap of the two counter-propagating generation
processes. A broad optical bandwidth corresponds to a short
characteristic temporal scale, making even small polarization-dependent
delays relevant when they remain far below the timing resolution of the
detectors.

Here we demonstrate an all-fiber type-0 Sagnac source based on a
pigtailed periodically poled lithium-niobate (PPLN) waveguide. Our aim
is to provide an affordable and practical source for fiber-based
quantum communication using commercially available components. The
fiber-connected design simplifies assembly and alignment while
providing direct interfaces to optical-fiber links
\cite{Vergyris2017}. The source generates non-degenerate telecom-band
photon pairs that are separated by wavelength within the WDM passbands, while
its broad emission spectrum offers a route to serving several user
pairs through passive wavelength multiplexing
\cite{Wengerowsky2018,Yadav2025}. The combination of broadband photon-pair generation and fiber compatibility may also be
relevant to quantum clock synchronization, which has been experimentally demonstrated over
fiber links using polarization-entangled photon pairs generated by SPDC \cite{Tang2023}.

In addition to reporting the source performance, we analyze the
reconstructed polarization density matrix to examine the balance
between population and coherence in the generated state. The measured
concurrence provides an upper bound on the improvement in Bell-state
fidelity attainable through local polarization transformations. The
density matrix then shows that the state remains predominantly within
the desired $\ket{HH}$--$\ket{VV}$ subspace, while the coherence
connecting these components is reduced from its maximum value. Finally,
we measure the evolution of the reconstructed state with pump power and
quantify the resulting brightness--quality trade-off.

% ============================================================
\section{Source design}
\label{sec:source}
% ============================================================

The source generates polarization-entangled photon pairs through type-0
spontaneous parametric down-conversion (SPDC) in a pigtailed periodically
poled lithium-niobate (PPLN) waveguide embedded in a fiber Sagnac
interferometer. The down-conversion process satisfies energy
conservation, $\omega_p=\omega_s+\omega_i$, and quasi-phase matching,
$k_p-k_s-k_i-2\pi/\Lambda\simeq0$, where $\omega_j$ and $k_j$ are
the angular frequency and wave number for $j=p,s,i$ (pump, signal, and
idler), and $\Lambda$ is the poling period. We label the
shorter-wavelength photon as signal and the longer-wavelength photon as idler.
In type-0 SPDC, the pump, signal, and idler fields are polarized along
the same crystal axis, providing access to the large nonlinear
coefficient of lithium niobate \cite{Tanzilli2002,Vergyris2017}.

\begin{figure}[t]
\centering
\figinclude[width=0.95\linewidth]{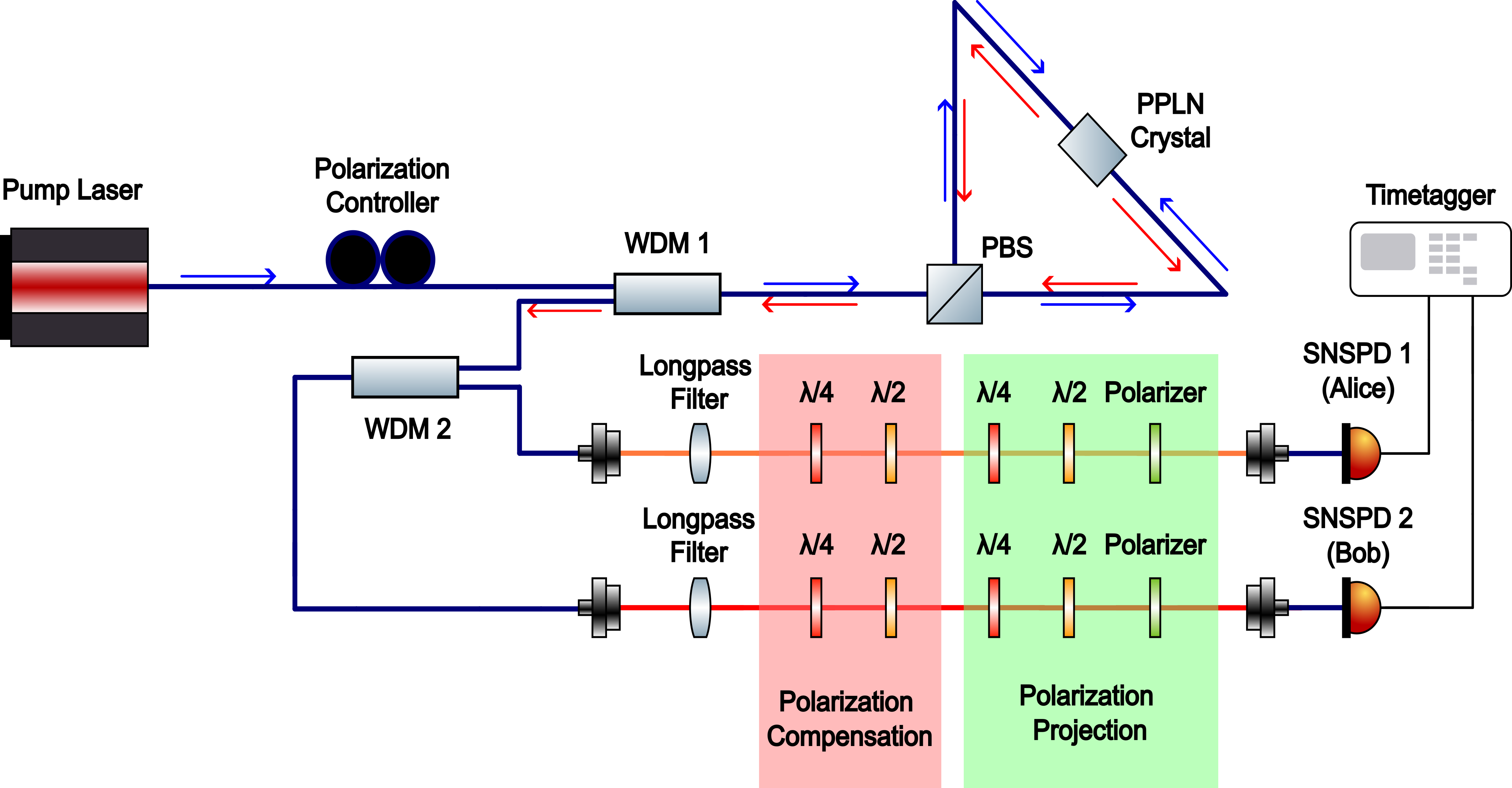}
\caption{Schematic of the all-fiber type-0 Sagnac source and
polarization-analysis setup. The fiber PBS launches the two pump
polarization components in opposite directions through the PPLN
waveguide. After recombination, the non-degenerate signal and idler
photons are separated by wavelength-division multiplexing, analyzed
independently, and detected by SNSPDs.}
\label{fig:setup}
\end{figure}

The experimental arrangement is shown in Fig.~\ref{fig:setup}. A
\SI{785}{\nano\meter} pump is injected through a fiber
wavelength-division multiplexer (WDM) and divided by a fiber polarizing beam
splitter (PBS) into clockwise and counter-clockwise components. A
\SI{90}{\degree} polarization transformation within the loop ensures
that both counter-propagating pump components address the phase-matched
polarization axis of the type-0 waveguide. The photon-pair amplitudes
generated in the two directions return to the PBS and recombine to form
the two polarization components of the entangled state. A second
WDM separates the non-degenerate signal and idler photons within its
passbands. They are then analyzed using
polarization optics and detected by superconducting nanowire
single-photon detectors (SNSPDs). The all-fiber designation refers to
the generation loop; the polarization analyzers include free-space optics.

For balanced pumping and indistinguishable clockwise and
counter-clockwise generation processes, the polarization state is
\begin{equation}
\ket{\Phi_\phi}
=
\frac{\ket{HH}+e^{i\phi}\ket{VV}}{\sqrt{2}},
\label{eq:source}
\end{equation}
where $\phi$ is the relative phase between the two Sagnac amplitudes. A
fixed value of $\phi$ determines which maximally entangled state is
produced but does not itself reduce the entanglement. For example,
$\phi=\pi$ gives the target state
$\ket{\Phi^-}=(\ket{HH}-\ket{VV})/\sqrt{2}$. Polarization coherence is
instead reduced when an unobserved degree of freedom, such as the
spectral--temporal mode, carries information that distinguishes the two
generation processes. This distinction is particularly important for
the broadband source considered here and is examined below using the
reconstructed polarization state.

The source is assembled from commercially available components,
including a pigtailed nonlinear waveguide, a fiber PBS, and WDMs.
Fiber connections eliminate the need to align free-space coupling
stages within the generation loop and simplify assembly, although
polarization and relative-delay adjustments remain important
\cite{Vergyris2017}. The aim of keeping costs low applies to the source
hardware; the SNSPDs and their cryogenic system used for
characterization are separate.

The measured state quality also has a direct connection to quantum
communication. For example, using the reconstructed two-qubit state as
a resource in a target-matched ideal qubit-teleportation protocol gives
an average fidelity over uniformly distributed pure input qubits
$\bar{f}_{\mathrm{tel}}=(2F_{\Phi^-}+1)/3=0.900$,
above the classical limit of $2/3$ \cite{Horodecki1999}. This value is
a theoretical benchmark based on the measured Bell-state fidelity,
rather than the result of a teleportation experiment.

The telecom wavelengths and fiber interfaces permit direct connection
to standard single-mode fiber links and passive wavelength-routing
components. Off-the-shelf multichannel WDMs with suitable passbands
could divide the broadband emission into several correlated
signal--idler channel pairs and route them to different users, allowing
the source to serve as the central node of an entanglement-distribution
network \cite{Wengerowsky2018,Yadav2025}. The selected channel pairs
must satisfy energy conservation and cover both emission bands. Such a
network extension would also require polarization entanglement to be
preserved within each channel pair.

% ============================================================
\section{Experimental characterization}
\label{sec:characterization}
% ============================================================

\begin{figure}[htbp]
\centering
\figinclude[width=\linewidth]{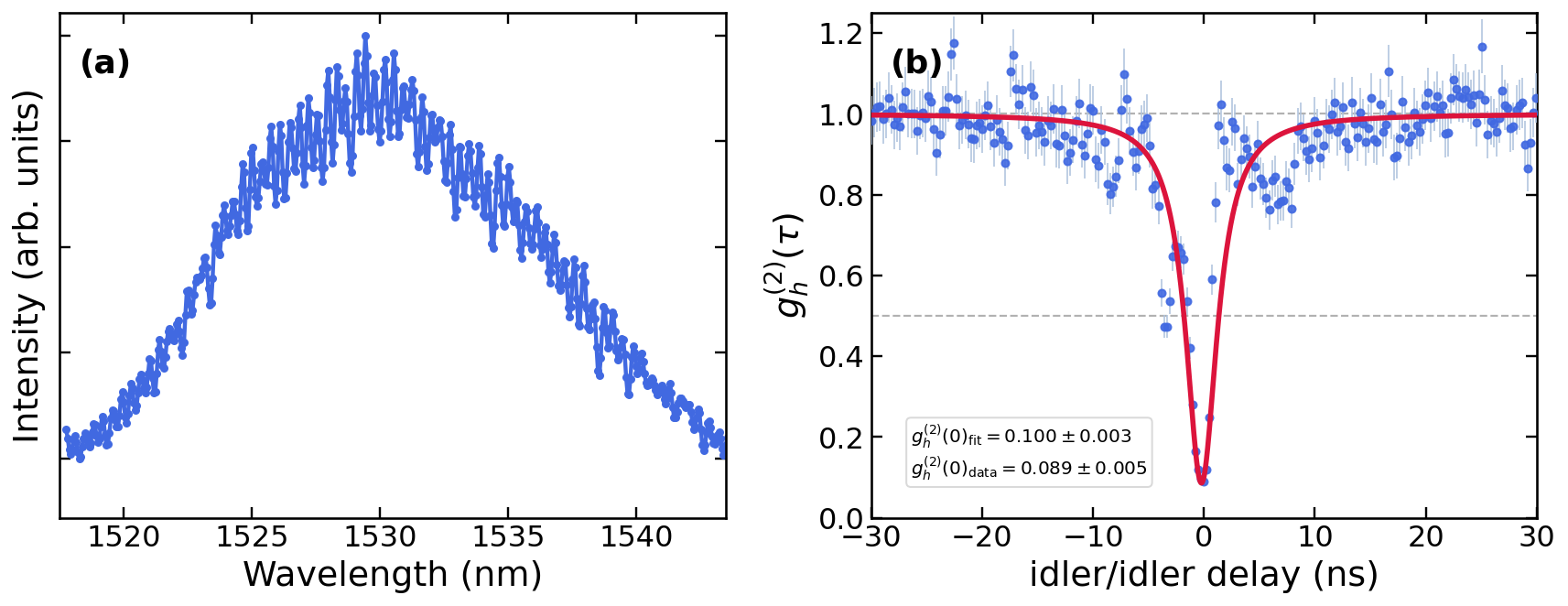}
\caption{Basic source characterization. (a) Measured lower-wavelength
SPDC sideband; the conjugate long-wavelength band lies outside the
available spectrometer range. (b) Heralded second-order correlation versus delay, sampled in
\SI{256}{\pico\second} steps. Error bars show propagated counting
uncertainties; the dashed line marks unity. No fit or additional
far-delay renormalization is applied.}
\label{fig:spdc_char}
\end{figure}

The basic source characterization is summarized in
Fig.~\ref{fig:spdc_char}. The lower-wavelength SPDC band,
shown in Fig.~\ref{fig:spdc_char}(a), has substantial emission approximately from
\SI{1520}{\nano\meter} to \SI{1542}{\nano\meter}, with a maximum near
\SI{1530}{\nano\meter}. The conjugate long-wavelength sideband lies
outside the range of the available spectrometer. For a pump wavelength
of \SI{785}{\nano\meter}, the degenerate wavelength is approximately
\SI{1570}{\nano\meter}, and energy conservation places the conjugate
photons on the opposite side of this wavelength. The resulting
non-degenerate spectrum allows the signal and idler photons to be
separated by wavelength-division multiplexing within the selected passbands.
For example, signal photons near \SI{1530}{\nano\meter} have conjugate
idler wavelengths near \SI{1612}{\nano\meter}.

The heralded second-order correlation was measured by using one photon
as a herald and routing the other through a $50{:}50$ fiber coupler to
two detectors. The zero-delay count bin gives
$g_H^{(2)}(0)=0.092\pm0.005$, consistent with a suppressed multiphoton
contribution to the heralded field. We use the count estimator
$g_H^{(2)}=N_HN_{Hab}/(N_{Ha}N_{Hb})$, where $N_H$ is the number of
heralds, $N_{Ha}$ and $N_{Hb}$ are herald--detector coincidences, and
$N_{Hab}$ is the triple-coincidence count \cite{Fasel2004}.
The quoted uncertainty uses independent-Poisson error propagation and
does not include timing-window or normalization systematics.

\begin{figure}[htbp]
\centering
\figinclude[width=0.70\textwidth]{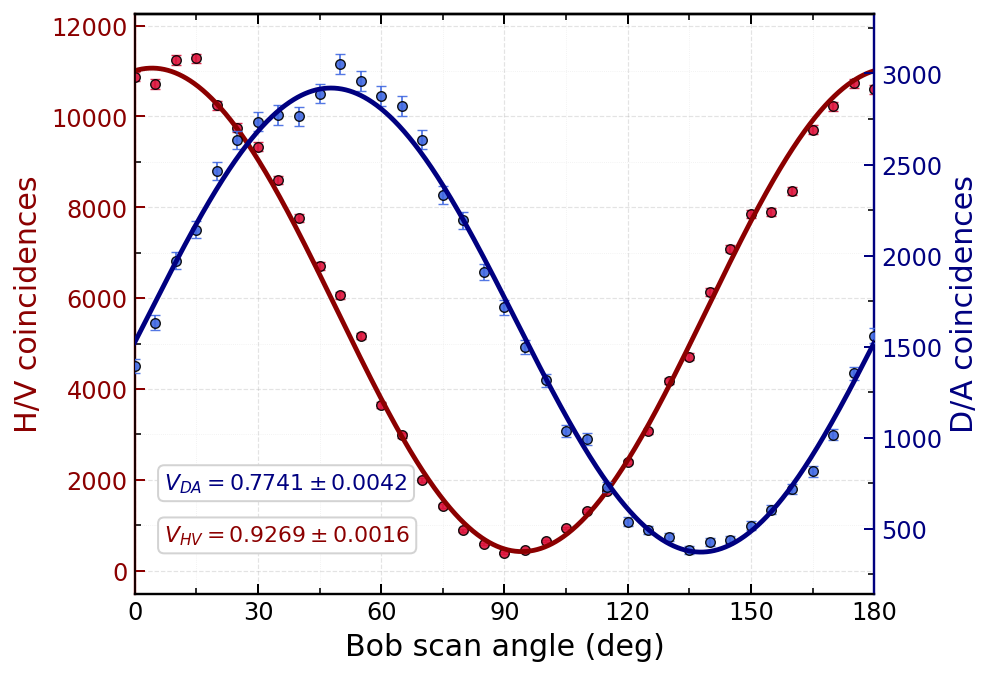}
\caption{Measured polarization-correlation fringes in the $H/V$ (red,
left axis) and $D/A$ (blue, right axis) bases. Error bars show central
Poisson confidence intervals at a nominal 68.27\% confidence level.
Solid lines are sinusoidal fits obtained by Poisson maximum likelihood.
Visibilities are calculated from the fitted amplitude-to-offset ratio;
the quoted uncertainties represent one standard deviation estimated
from 2,000 parametric Poisson-bootstrap resamples and refits. Each
point contains \SI{5}{\second} of acquisition. The fit has a
\SI{180}{\degree} period in the recorded scan coordinate; this
coordinate is not identified with a calibrated physical half-wave-plate
angle.}
\label{fig:correlation_curves}
\end{figure}

Polarization correlations were measured in the $H/V$ and $D/A$ bases,
with $\ket{D}=(\ket{H}+\ket{V})/\sqrt{2}$ and
$\ket{A}=(\ket{H}-\ket{V})/\sqrt{2}$. Poisson maximum-likelihood fits give
$\mathcal{V}_{HV}=0.9269\pm0.0016$ and
$\mathcal{V}_{DA}=0.7741\pm0.0042$, with counting-only bootstrap
uncertainties. The measured fringes and corresponding sinusoidal fits
are shown in Fig.~\ref{fig:correlation_curves}. The contrast in the two
scans probes different polarization correlations; the diagonal-basis
contrast depends on both the phase and magnitude of the $HH$--$VV$
coherence and is not by itself a measure of decoherence.
The Poisson deviances per degree of freedom are $16.4$ and $3.2$,
respectively, indicating variation beyond the fitted counting-noise
model. The quoted uncertainties therefore do not describe the full
experimental reproducibility.

Full two-qubit polarization tomography was performed using the sixteen
product projections formed from $H,V,D,R$ on each arm, with
$\ket{R}=(\ket{H}+i\ket{V})/\sqrt{2}$, followed by Poisson
maximum-likelihood reconstruction constrained to a positive-semidefinite,
unit-trace density matrix \cite{James2001,Altepeter2005}. At a pump power of
\SI{0.5}{\milli\watt}, with \SI{5}{\second} per projection, the
reconstructed state has a Bell-state fidelity $F_{\Phi^-}=0.850$,
purity $\mathcal{P}=0.836$, and concurrence $C=0.782$
\cite{Wootters1998}, where
$F_{\Phi^-}=\bra{\Phi^-}\rho\ket{\Phi^-}$ and
$\mathcal{P}=\Tr(\rho^2)$. Their counting-only standard deviations,
obtained from 1,000 parametric Poisson-bootstrap reconstructions, are
$0.0064$, $0.0109$, and $0.0149$, respectively. No accidental-coincidence
subtraction is applied to the reconstruction.

A separate CHSH acquisition gives $S=2.3301\pm0.0075$, exceeding the
local-realistic bound $|S|\leq2$ by $44.1$ counting-statistical standard
deviations, calculated from the unrounded values \cite{Clauser1969}.
Because the measurement relies on post-selected coincidences, the CHSH
value is used here to characterize the measured polarization
correlations rather than to claim a loophole-free Bell test. The
complete reconstructed density matrix is analyzed in the following
section to identify the features responsible for the departure of the
measured state from the ideal Bell state.

\section{Analysis of the reconstructed state}
\label{sec:state_analysis}
% ============================================================

The measured fidelity can first be compared with the amount of
entanglement in the reconstructed state. Concurrence is invariant under
local unitary transformations, including ideal polarization rotations
implemented using wave plates or polarization controllers. For a
two-qubit state, the maximum fidelity with a maximally entangled state
attainable through local unitary transformations is bounded by
$F_{\max}\leq(1+C)/2$ \cite{Verstraete2002}. For the measured
concurrence $C=0.782$, this gives $F_{\max}\leq0.891$, compared with
the measured value $F_{\Phi^-}=0.850$. Local polarization
transformations can therefore improve the Bell-state fidelity only
within a limited range. The reconstructed density matrix can be used
to identify the main contributions to the remaining infidelity.

In the ordered basis
$\{\ket{HH},\ket{HV},\ket{VH},\ket{VV}\}$,
the state at \SI{0.5}{\milli\watt} pump power is
\begin{equation}
\rho \simeq
\resizebox{0.88\textwidth}{!}{$
\begin{pmatrix}
0.359755 & 0.001926+0.020079i & -0.009355-0.017232i & -0.366691+0.167257i \\
0.001926-0.020079i & 0.017978 & 0.003414+0.013045i & -0.000618+0.060576i \\
-0.009355+0.017232i & 0.003414-0.013045i & 0.015111 & 0.040640-0.013846i \\
-0.366691-0.167257i & -0.000618-0.060576i & 0.040640+0.013846i & 0.607156
\end{pmatrix}
$}.
\label{eq:measured_density_matrix}
\end{equation}
All derived quantities use the unrounded reconstruction.

The state is dominated by the $\ket{HH}$ and $\ket{VV}$ populations,
with $\rho_{HH,HH}\simeq0.35975$ and
$\rho_{VV,VV}\simeq0.60716$, together accounting for approximately
$96.7\%$ of the total population. Only approximately $3.3\%$ lies in
the $\ket{HV}$ and $\ket{VH}$ components. The dominant coherence
connecting the desired components has magnitude
$|\rho_{HH,VV}|\simeq0.40303$.

For the target state $\ket{\Phi^-}$, these matrix elements reproduce
the measured fidelity $F_{\Phi^-}\simeq0.850$. The density matrix
therefore shows that the main Bell-state components are strongly
populated, while the fidelity is additionally limited by their
imbalance, the small population leakage into $\ket{HV}$ and
$\ket{VH}$, and the magnitude and relative phase of the coherence
between $\ket{HH}$ and $\ket{VV}$.

To account for the unequal $\ket{HH}$ and $\ket{VV}$ populations, the
coherence can be normalized to its maximum value allowed by positivity,
$|\rho_{HH,VV}|\leq\sqrt{\rho_{HH,HH}\rho_{VV,VV}}$. This gives
\begin{equation}
|\Gamma_{\mathrm{eff}}|
=
\frac{|\rho_{HH,VV}|}
{\sqrt{\rho_{HH,HH}\rho_{VV,VV}}}
\simeq0.862.
\label{eq:gamma_eff}
\end{equation}
The measured $HH$--$VV$ coherence is therefore approximately $86\%$
of the maximum compatible with the observed populations. This quantity
characterizes the reconstructed polarization state but does not by
itself identify the physical mechanism responsible for the reduction.

One possible contribution to the reduced coherence is
spectral--temporal mismatch between the clockwise and counter-clockwise
generation processes. A differential delay, for example from
birefringence within the loop, introduces a frequency-dependent
relative phase between the two generation amplitudes. Tracing over
unresolved spectral--temporal degrees of freedom can then reduce the
observed polarization coherence \cite{Antonelli2011}. The delay need
not be resolved by the detectors: collecting both alternatives within
the same coincidence window does not restore their quantum-mode overlap.

For scale, a spectral FWHM of
$\Delta\lambda=\SI{12}{\nano\meter}$ around
$\lambda=\SI{1530}{\nano\meter}$ corresponds to a frequency bandwidth
of approximately \SI{1.54}{\tera\hertz}, whose inverse is approximately
\SI{0.65}{\pico\second}. This provides a characteristic temporal scale
over which spectral--temporal distinguishability can become relevant.

As an order-of-magnitude illustration, assuming identical Gaussian
spectral-intensity profiles for the two generation alternatives, the
measured normalized coherence
$|\Gamma_{\mathrm{eff}}|\simeq0.862$ would correspond to an effective
differential delay of approximately \SI{0.13}{\pico\second}. For an
assumed PM-fiber differential group delay per unit length of
\SI{1.5}{\pico\second\per\meter}, this is equivalent to approximately
\SI{8.9}{\centi\meter} of fiber.

This estimate should not be interpreted as a measurement of the actual
fiber imbalance. The relation between spectral--temporal mismatch and
the observed polarization coherence depends on the detailed spectral
profile and on the filtering in both arms. In the narrowband-pump
limit, the relevant effective delay may take the form
$\Delta\tau=\delta\tau_s-\delta\tau_i$, where $\delta\tau_s$ and
$\delta\tau_i$ are the relative delays between the two generation
alternatives in the signal and idler arms \cite{Antonelli2011}.
The estimate therefore only illustrates the sensitivity of the
broadband source to sub-picosecond differential delays, without
identifying such a delay as the origin of the measured coherence loss.

% ============================================================
\section{Pump-power dependence}
\label{sec:power}
% ============================================================

\begin{figure}[H]
\centering
\figinclude[width=0.9\textwidth]{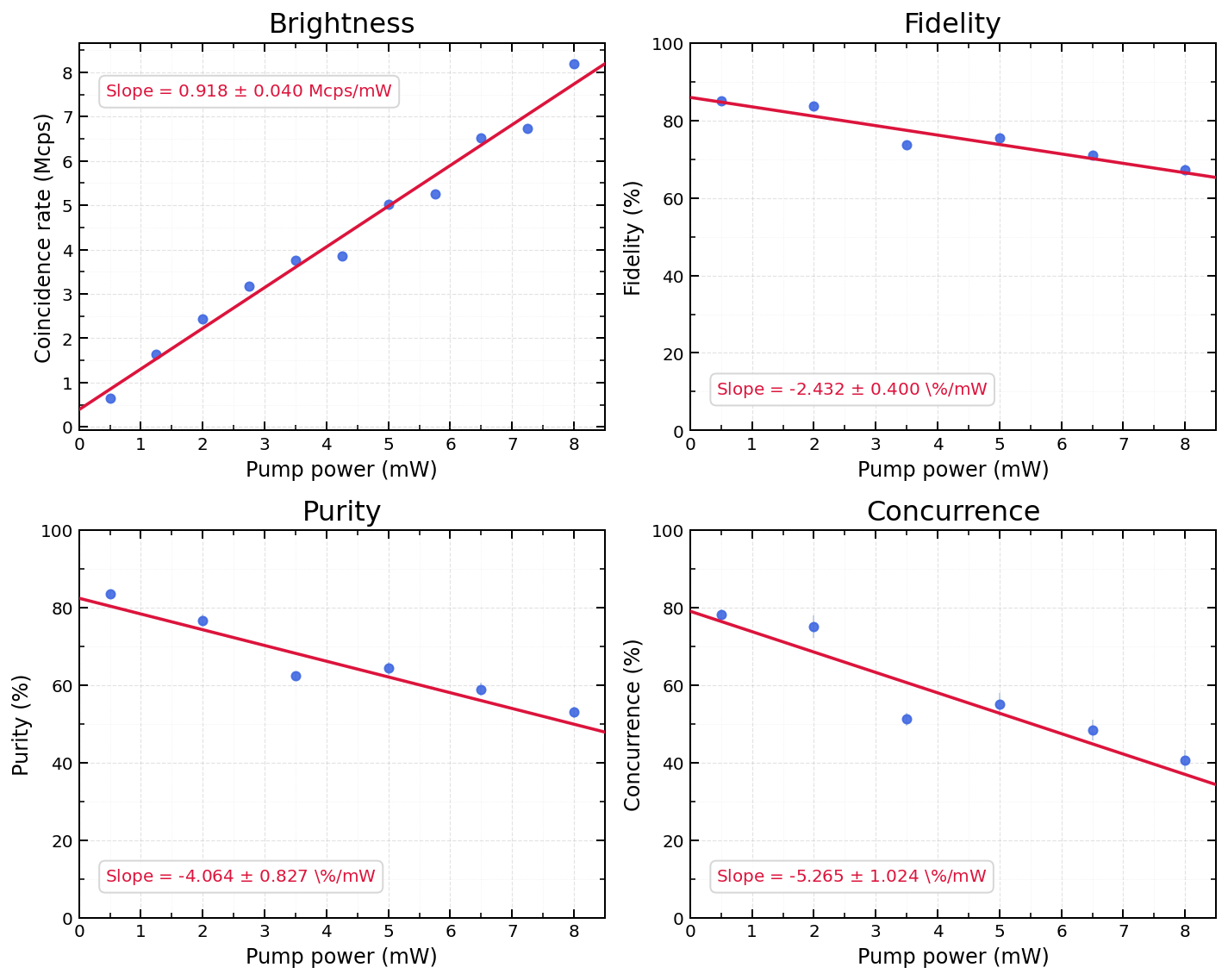}
\caption{Pump-power dependence of (a) the corrected pair rate,
(b) Bell-state fidelity, (c) purity, and (d) concurrence. The pair
rate increases approximately linearly with pump power, while the
state quality decreases overall. Error bars in (b)--(d) represent
one standard deviation of the reconstructed metrics. Solid lines
are linear fits. State-metric slopes are expressed in percentage
points per milliwatt.}
\label{fig:power_dependence}
\end{figure}

The corrected pair rate and reconstructed polarization state were
measured as functions of pump power $P$, as shown in
Fig.~\ref{fig:power_dependence}. The corrected pair rate increases
approximately linearly over \SIrange{0.5}{8}{\milli\watt}, with a
fitted slope of \SI{0.918(40)}{\mega\cps\per\milli\watt}. The
fidelity, purity, and concurrence decrease overall, with respective
fitted slopes of
\SI{-2.432(400)}{\percent\per\milli\watt},
\SI{-4.064(827)}{\percent\per\milli\watt}, and
\SI{-5.265(1024)}{\percent\per\milli\watt}.
The state-metric slopes are expressed in percentage points per
milliwatt. The linear fits include a free intercept.

The observed trend is consistent with increasing contributions from
higher-order pair generation and accidental coincidences. These measurements quantify the
brightness--quality trade-off without attributing it to a single
microscopic mechanism.

% ============================================================
\section{Conclusion}
% ============================================================

We have demonstrated an all-fiber source of non-degenerate telecom-band
polarization-entangled photon pairs based on type-0 SPDC in a pigtailed
PPLN waveguide embedded in a Sagnac interferometer. At
\SI{0.5}{\milli\watt}, tomography gives $F_{\Phi^-}=0.850$,
$\mathcal{P}=0.836$, and $C=0.782$. The CHSH acquisition yields
$S=2.3301\pm0.0075$.

Analysis of the reconstructed state shows that approximately $96.7\%$
of the population remains within the desired
$\ket{HH}$--$\ket{VV}$ subspace, while the normalized coherence between
these components is $|\Gamma_{\mathrm{eff}}|\simeq0.86$. The measured
concurrence limits the Bell-state fidelity attainable through local
polarization transformations to $F_{\max}\leq0.891$. The phase of the
$HH$--$VV$ coherence indicates that local phase compensation could
increase the fidelity to approximately $0.8865$, whereas the remaining
infidelity reflects population imbalance, leakage, and incomplete
coherence that cannot be removed through phase compensation alone.

The broadband spectrum makes spectral--temporal overlap a relevant
design consideration, although the present measurements do not isolate
its contribution to the observed coherence loss. The pump-power
dependence also shows the need to balance coincidence rate against
polarization-state quality.

The main practical outcome is a fiber-connected source assembled from
off-the-shelf components, with affordability and ease of assembly and
alignment as central design objectives. Its measured state quality
supports its use as a resource for quantum communication, as
illustrated by the ideal teleportation benchmark. Compatibility with
standard optical-fiber links and the possibility of adding matched
multichannel WDMs provide a route from the present source to a
multiuser entanglement-distribution platform.

% ============================================================
\section*{Acknowledgments}
% ============================================================

The authors gratefully acknowledge support from the Knut and Alice
Wallenberg Foundation through the Wallenberg Centre for Quantum
Technology (WACQT).

% ============================================================
% References
% ============================================================

\end{document}